# FIRST RESULTS OF AUP Nb$_3$Sn QUADRUPOLE HORIZONTAL TESTS

M. Baldini†, G. Ambrosio, G. Apollinari, J. Blowers, R. Bossert, R. Carcagno, G. Chlachidze, J. DiMarco, S. Feher, S. Krave, V. Lombardo, L. Martin, C. Narug, T. H. Nicol, V. Nikolic, A. Nobrega, V. Marinozzi, C. Orozco, T. Page, S. Stoynev, T. Strauss, M. Turenne, D. Turrioni, A. Vouris, M. Yu, Fermi National Accelerator Laboratory, Batavia, Illinois 60510, USA

A. Baskys, D. Cheng, J. F. Croteau, P. Ferracin, L. Garcia Fajardo, E. Lee, A. Lin, M. Marchevsky, M. Naus, H. Pan, I. Pong, S. Prestemon, K. Ray, G. Sabbi, C. Sanabria, G. Vallone, X. Wang, Lawrence Berkeley National Laboratory, Berkeley CA 94720

K. Amm, M. Anerella, A. Ben Yahia, H. Hocker, P. Joshi, J. Muratore, J. Schmalzle, H. Song, P. Wanderer, Brookhaven National Laboratory, Upton, NY 11973

## Abstract

The Large Hadron Collider will soon undergo an upgrade to increase its luminosity by a factor of ~10 [1]. A crucial part of this upgrade will be replacement of the NbTi focusing magnets with Nb$_3$Sn magnets that achieve a ~50% increase in the field strength. This will be the first ever large-scale implementation of Nb$_3$Sn magnets in a particle accelerator. The High-Luminosity LHC Upgrade, HL-LHC is a CERN project with a world-wide collaboration. It is under construction and utilizes Nb$_3$Sn Magnets (named MQXF) as key ingredients to increase tenfold the integrated luminosity delivered to the CMS and ATLAS experiments in the next decade.

The HL-LHC AUP is the US effort to contribute approximately 50% of the low-beta focusing magnets and crab cavities for the HL-LHC.

This paper will present the program to fabricate the Nb$_3$Sn superconducting magnets. We are reporting the status of the HL-LHC AUP project present the results from horizontal tests of the first fully assembled cryo-assembly.

## INTRODUCTION

The reduction of the transverse beam size by approximately a factor of two in the interaction points will be achieved in HL-LHC through the installation of new inner triplet, low-beta cryo-assemblies. The interaction region magnet triplet consists of four elements: Q1, Q2a, Q2b and Q3. Q1/Q3 cryo-assembly contain two 4.2 m quadrupole magnets (MQXFA) whereas Q2a and Q2b consist of a single unit MQXFB ~7.5-m-long quadrupole magnet and high order multiple corrector magnets [2]. The Q2 elements are fabricated by CERN and the Q1/Q3 elements by US effort [3]. The only significant difference between MQXFA and MQXFB is the magnetic length. The Nb$_3$Sn quadrupole magnets operate in superfluid He at 1.9 K with a nominal field gradient of 132.2 T/m [4-7].

HL-LHC AUP is a collaboration among four US laboratories: Brookhaven National Laboratory (BNL), Florida State University (FSU), Lawrence Berkeley National Laboratory (LBNL) and Fermi national Accelerator Laboratory (FNAL). The superconductor strand procurement and test are Fermilab and FSU responsibilities. The cable production is being done by LBNL, the coils are being fabricated at BNL and FNAL, the magnets are being produced in LBNL then vertical tests are conducted at BNL. The Cold Masses containing two magnets were designed and are being fabricated by FNAL. The cryostat design and procurement of the cryostat parts are CERN responsibilities. The final Cryo-Assemblies (LQXFA/B) are tested at FNAL.

In this paper we provide an overview of the status of the production, and testing of MQXFA magnets and we will discuss the design, fabrication, and test of the first Cryo-Assembly that was completed at the beginning of fall 2023.

## THE MQXFA QUADRUPOLES

The cornerstones of the HL-LHC project are the low-beta quadrupole magnets (MQXF) [7] with an unprecedentedly large aperture (150 mm) and gradient (132.6 T/m). MQXF are the first magnets to use Nb$_3$Sn conductor in a particle accelerator, paving the way for the use of this material in future high-energy colliders. The main challenges facing these magnets include: 1) the electromagnetic forces they are subject to, which in the straight section and in the ends (1.15 MN) are four and six times higher, respectively, than in the LHC low-beta magnets, and 2) the energy density in the coils (78 mJ/mm$^3$), which is about double the energy density in the coils of LHC low-beta magnets. The design of those magnets has been the results of more than 15 years of research and development supported by direct R&D program LARP. Some details about design challenges can be found in [8, 9]. The Coil and Magnet parameters and requirements are reported in ref. [3].

### MQXFA fabrication process and status

As mentioned in the introduction, the MQXFA fabrication includes activities performed by several US

___________________

* Work supported by the U.S. Department of Energy, Office of Science, Office of High Energy Physics, through the US LHC Accelerator Upgrade Project, and by the High Luminosity LHC project at CERN.
† Corresponding author: mbaldini@fnal.gov

laboratories. The strand used is RRP 108/127 by Bruker-OST [10-12]. Strand procurement is managed by FNAL and strand QC verification processes is implemented and performed at NHMFL. Cables are fabricated by LBNL. By the end of April 2024, the strand procurement was complete as well as the cable fabrication. More than 50 Km of $Nb_3Sn$ Rutherford cable (113 cables, 470 m per coil) has been produced with a fabrication yield which is higher than 95%.

Coil parts are procured by FNAL and sent to either FNAL or BNL for coil fabrication [13]. QC verification of those parts is also performed at FNAL. By the end of April 2024 coil fabrication at BNL is complete and only one coil is left to be built at FNAL. A total number of 110 coils have been produced by both laboratories with a fabrication yield of 89%.

Magnet assembly is performed at LBNL with bladder and key technology and 60% of the production has been completed. [14, 15]. This is the first time using this technology for an accelerator magnet. The MQXFA quadrupoles are tested individually at the vertical magnet test facility of the Superconducting Magnet Division at BNL in superfluid He at 1.9 K and 1 bar [16]. This test aims at verifying that each magnet meets the HL-LHC acceptance criteria [17, 18] before it is used in an LMQXFA cold mass. Currently, 13 magnets underwent vertical test with 10 of them passing the Acceptance Criteria. Fig. 1 displays the quench training curves of the 10 magnets meeting requirements.

understood and included Covid restrictions on personnel. The investigation was performed in collaboration with CERN. Results are reported in detail in Ref [19, 20].

MQXFA05 magnet underwent an endurance test after it passed all Acceptance Criteria and was stored for over a year. It demonstrates that MQXFA magnets can sustain 3 thermal cycles, 79 power cycle and more than 50 quenches. The resilience of the MQXFA magnets was also demonstrated by MQXFA11 that met requirements during vertical testing after the truck transporting it was heavily damaged in a highway crash.

Recent tests performed on short MQXF magnets (MQXFS) at CERN [21] and lesson learned from MQXFA13 test analysis suggested that increasing preload can be beneficial for magnet performance. A maximum stress of -110 MPa in the coil at room temperature was set at project start based on experience from LARP model magnets [22]. This specification limit has been recently increased to -120 MPa allowing higher preload in the last three magnets (MQXFA14b, MQXFA7b and MQXFA15). Quench training curves suggest that training is significantly reduced with the higher prestress (see Fig. 2). This result is very significant because the number of training quenches is a crucial parameter for the construction of future accelerator machines.

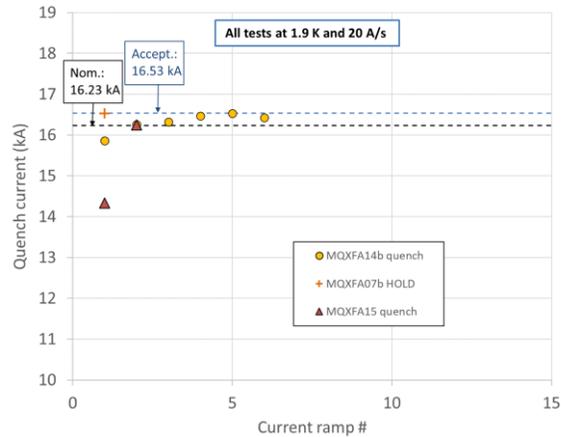

Figure 2: Quench history of MQXFA magnets 14b, 7b, and 15 during vertical test at BNL. The number of training quenches is significantly reduced.

## THE LQXFA/B CRYO-ASSEMBLY

The Cold Mass design [23, 24] is a Stainless-steel pressure vessel made from two half shells and two end covers welded together. Since the Cold Mass contains two 4.2m magnetic-length MQXFA $Nb_3Sn$ quadrupoles the Cold Mass design takes into account the following: the alignment of the two MQXFA magnets, the bus work that connects the two magnets and provides the through bus, the heat exchangers, the beam tube, and the magnet instrumentation wiring from the magnets all the way to the cryostat ports.

Cold Mass production [23- 26] starts with the inspection (electrical and mechanical integrity) of the two previously

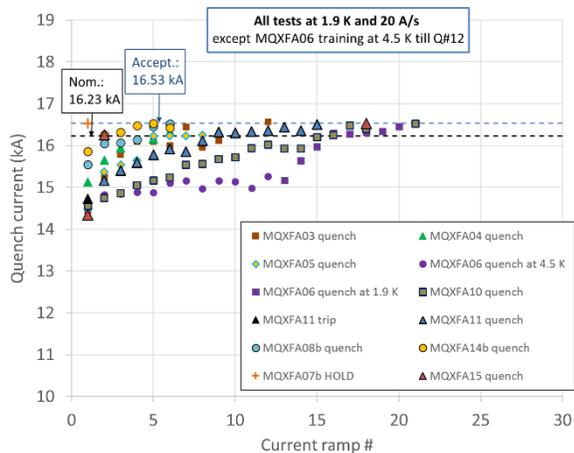

Figure 1: Quench history of MQXFA magnets 03, 04, 05, 06, 10, 11, 8b, 14b, 7b and 15 during vertical test at BNL (only training quenches are shown). MQXFA06 was trained at 4.5 K until quench #12.

The two main requirements to be verified are holding current at acceptance (16530A) current and demonstrating a good training memory (no more than 3 quenches) after a thermal cycle [18]. The MQXFA07, MQXFA08 and MQXFA13 were disassembled, because were limited in performance and the liming coil was replaced with a conforming coil. Two of these three magnets were then successfully tested. MQXFA13b will be tested again over the next couple of months. The root causes for the limiting performances of MQXFA07 and MQXFA08 have been well cold tested magnets. The Cryo-Assembly design contains a Cold Mass surrounded with pipes for the thermal shield, beam screen, quench recovery, bus work, and superfluid pumping. The cryostat tooling was designed provided to FNAL via CERN. Cryostat activities are reported in Ref. [27].

The first pre-series US-AUP Cold Mass (LMQXFA-01) production was completed in early fall of 2022. Cryostat work including the final pressure and leak test was completed in January 2023. The first Cryo-Assembly (LQXFA/B-01) was tested in the newly upgraded horizontal test facility at Fermilab. The commissioning of the test facility with LQXFA/B-01 test and the cold tests were completed at the beginning of fall 2023 [23].

### Fermilab Horizontal test stand

Once the Cryo-Assembly is completed is moved to the recently upgraded horizontal test facility [28] also known as Stand 4 of the Fermilab magnet test facility. A 30kA power supply system can power the magnets at the horizontal test stand using water-cooled solid bus bars and water-cooled flexible buses. A picture of the LQXFA/B-01 cryo-assembly on the horizontal test stand is shown in Fig. 3.

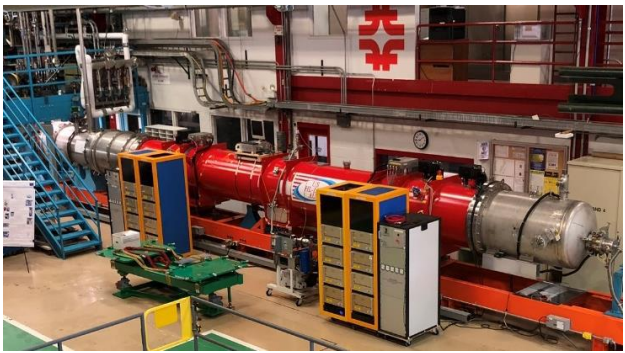

Figure 3: LQXFA/B-01 cryo-assembly on the horizontal test stands.

The test facility was upgraded and refurbished to meet the High Luminosity LHC (HL-LHC) cryo-assembly design and test requirements [28]. An important addition to the stand is the new Adapter box designed and built as an interface between the existing cryogenic feed box and the new HL-LHC cryo-assemblies. It consists of a vacuum shell and thermal shield, and helium piping.

The Quench Protection and Monitoring (QPM) system is designed for a high level of reliability based on two independent and fully redundant Tier-1 and Tier-2 systems (primary FPGA based Digital Quench Detection (DQD) system and secondary Analog Quench Detection (AQD) system). Tier 3 is the Quench controls and Data Management system. The upgraded test facility was successfully tested without a superconducting magnet (zero magnet test) being attached to it [28]. Therefore, the test of the first Cryo-Assembly and the commissioning of the test facility were done at the same time making the test last longer than it would have been with a fully commissioned facility.

### Horizontal test results for the first LQXFA/B cryo-assembly

The test of the first LQXFA/B cryo-assembly was successfully completed in the fall 2023 [31]. The LQXFA/B-01 cryo-assembly was assembled with MQXFA03 and MQXFA04 magnets.

The main requirements for testing are listed below:
- The nominal operating current is $I_{nom}$=16230 A, and the acceptance current is $I_{nom}$ + 300 A= 16530 A.
- Each MQXFA magnet shall be able to withstand a maximum temperature gradient of 50 K during a controlled warm-up or cool-down.
- The MQXFA magnets shall be capable of operating at any ramp rate within ±30 A/s and shall not quench while ramping down at 150 A/s from the nominal operating current.
- After a thermal cycle (TC) to room temperature, MQXFA magnets shall attain the nominal operating current with no more than 3 quenches.
- Splice resistance must be less than 1.0 n$\Omega$ at 1.9 K.
- LMQXFA cold mass shall be capable of continuous steady-state operation at nominal current in pressurized static superfluid helium (HeII) at 1.3 bar and at a temperature of 1.9 K. The requirement for LQXFA/B cryo-assemblies is to hold magnets at nominal current at 1.9 K for 300 minutes.

Two quench protection elements were employed during the horizontal test of the first Cryo-assembly: Heater Firing Units (HFU) applied to the outer layer coil quench heaters (eight per magnet) [29], and Coupling-Loss Induced Quench (CLIQ) units [30]. The HFU voltage and capacitance settings are fixed at 600 V and 7.05 mF respectively. The corresponding CLIQ settings are 50 mF and 600 V. Both protection elements are initiated at quench detection with no delay.

**Cooldown** The cooldown of LQXFA/B-01 was performed in three steps – down to liquid nitrogen (LN), below LN temperatures with a mixture of 80 K He gas and liquid, and final cool down by liquid helium. The temperature difference between the cryo-assembly ends was kept below 80 K. This constrain is sufficient to verify that a maximum temperature gradient of 50 K is present in each magnet. A failure of the coil to heater high voltage test was observed during the electrical checkout at 4.2 K. High voltage tests at cold on future magnets will be performed at 1.9 K The failure does not limit the effectiveness of the quench protection system and the cryo-assembly can still be safely protected. The hi-pot failure was treated as a non-conformity.

**First cooldown: Quench Training Results** The quench performance at 1.9 K during the first thermal cycle is shown in Fig. 4. The two MQXFA magnets in the LQXFA/B-01 cryo-assembly reached acceptance current without any spontaneous quenches. Only one spontaneous quench was observed during the first thermal-cycle at 16386 A. However, because of liquid level instabilities, two quenches (first two cross symbols in Fig. 4) were observed in the superconducting leads outside the cryo-assembly.

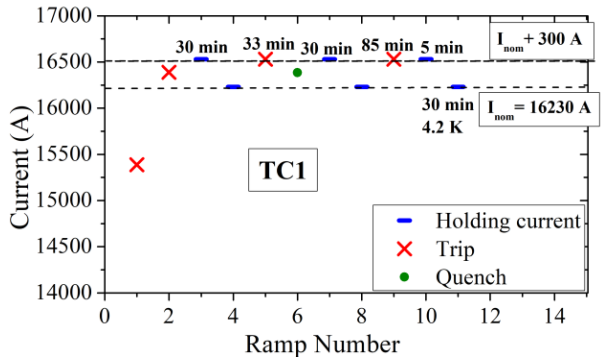

Figure 4: Quench performance during the first thermal cycle. Cross symbols mark Cryo-assembly trips due to cryogenic instabilities and test facilities threshold settings.

**Second cooldown: Quench Training Results** The cryo-assembly quench performance in the second thermal cycle is shown in Fig. 5. One spontaneous quench was observed at 16525 A in the MQXFA03 magnet. Magnetic measurements were performed at the nominal current at 1.9 K. Detailed results are presented in [32].

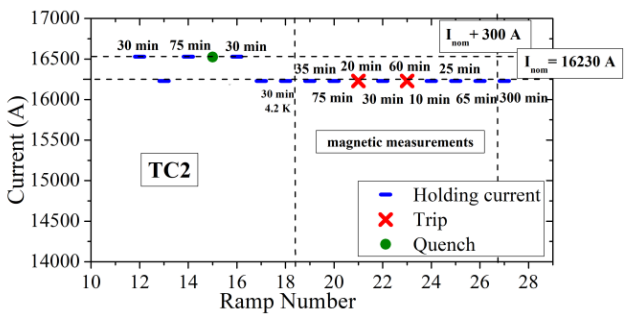

The cause of instability in the liquid helium lines was understood and the problem was resolved at the very end of the second thermal cycle. The five-hour holding test at the nominal current of 16230 A was successfully completed. The first Cryo-assembly was then shipped to CERN where the acceptance test is currently undergoing.

Figure 5: Quench performance during the second thermal cycle. Cross symbols mark trips due to cryogenic instabilities. Magnetic measurements were performed holding current at 16.23 kA.

## CONCLUSION

This paper presents the status of the US-contribution to the HL-LHC. Strand procurement and cable fabrication are complete. Coil fabrication is 99% complete. The magnet assembly has reached 60% of production. Thirteen magnets were vertically tested and 10 of them met the acceptance criteria. Two more magnets met requirements after replacing the limiting coils. The third one will be tested soon. The first LQXFA/B cryo-assembly was successfully fabricated and tested. All acceptance criteria were met.

The second Cryo-assembly is currently being completed and is under test at FNAL. The third cold mass is completed and is being installed in the cryostat. The stainless-steel shell is being installed on 4$^{th}$ cold mass after magnet alignment.

## REFERENCES

[1] O. Brüning and L. Rossi, The High Luminosity Large Hadron Collider, Advanced Series on Directions in High Energy Physics, vol. 24, Oct. 2015

[2] E. Todesco et al., "A first baseline for the magnets in the high luminosity LHC insertion regions", IEEE Trans. Appl. Supercond., vol. 24 no. 3, pp. 4003305, Jun 2014. doi:10.1109/TASC.2013.2288603

[3] G. Apollinari et al., "US Contribution to the High Luminosity LHC Upgrade: Focusing Quadrupoles and Crab Cavities", J. Phys.: Conf. Ser., vol. 1350, p. 012005, May 2019. doi:10.1088/1742 6596/1/1350/012005

[4] E. Todesco et al., "Design studies for the low-Beta quadrupoles for the LHC luminosity upgrade, IEEE Trans. Appl. Supercond., vol. 23 no. 3, pp. 4002405, Jun. 2013.
doi:10.1109/TASC.2013.2248314

[5] P. Ferracin et al., "Magnet design of the 150 mm aperture low- quadrupoles for the High Luminosity LHC", IEEE Trans. Appl. Supercond., vol. 24 no. 3, pp. 4002306, Jun. 2014. doi:10.1109/TASC.2013.2284970

[6] P. Ferracin et al., "Development of MQXF: The Nb$_3$Sn low- quadrupole for the HL-LHC", IEEE Trans. Appl. Supercond., vol. 26, no. 4, pp. 4000207, Jun. 2016. doi:10.1109/TASC.2015.2510508.

[7] E. Todesco et al., "Review of the HL_LHC interaction region magnets towards series production," *Superconducting Sci. Technol.*, vol. 34, 2021, Art. no. 054501.

[8] G. Ambrosio and P. Ferracin, "Large-aperture high-field Nb$_3$Sn quadrupole magnets for HiLumi," in *The Future of the Large Hadron Collider: A Super-Accelerator With Multiple Possible Lives*, O. Bruning, M. Klein, L. Rossi, and P. Spagnolo, Eds., Singapore: World Scientific (2023); https://doi.org/10.1142/9789811280184_0007

[9] P. Ferracin et al., "The HL-LHC low-beta quadrupole magnet MQXF: From short model to long prototype," *IEEE Trans. Appl. Supercond.*, vol. 29, no. 5, Aug. 2019, Art. no. 4001309.

[10] L. D. Cooley et al., "Conductor Specification and Validation for High Luminosity LHC Quadrupole Magnets", IEEE Trans. Appl. Supercond., vol. 27 no. 4, pp. 1-5, Jun. 2017.
doi:10.1109/TASC.2017.2648738.

[11] L. Cooley et al., "Challenges and opportunities to assure future manufacturing of magnet conductors," *arXiv:2208.12379*.

[12] A. Ballarino, "Nb$_3$Sn conductor at CERN for HL-LHC and beyond," *IEEE Trans. Appl. Supercond.*, to be published

[13] G. Ambrosio, "Nb$_3$Sn High Field Magnets for the High Luminosity LHC Upgrade Project," in *IEEE Transactions on Applied Superconductivity*, vol. 25, no. 3, pp. 1-7, June 2015, Art no. 4002107, doi: 10.1109/TASC.2014.2367024.

[14] D. W. Cheng *et al.*, "Fabrication and Assembly Performance of the First 4.2 m MQXFA Magnet and Mechanical Model for the Hi-Lumi LHC Upgrade," in *IEEE Transactions on Applied Superconductivity*, vol.

28, no. 3, pp. 1-7, April 2018, Art no. 4006207, doi: 10.1109/TASC.2018.2799563.

[15] D. W. Cheng *et al*., "Fabrication and Assembly Performance of the First 4.2 m MQXFA Magnet and Mechanical Model for the Hi-Lumi LHC Upgrade," in *IEEE Transactions on Applied Superconductivity*, vol. 28, no. 3, pp. 1-7, April 2018, Art no. 4006207, doi: 10.1109/TASC.2018.2799563.

[16] J. Muratore, M. Anerella, P. Joshi, P. Kovach, A. Marone, and P. Wanderer, "Design and fabrication of the 1.9 K magnet test facility at BNL, and test of the first 4-m-long MQXF coil," *IEEE Trans. Appl. Supercond.*, vol. 28, no. 3, Apr. 2018, Art no. 9500104.

[17] J. Muratore et al., "Test results of the first pre-series quadrupole magnets for the LHC Hi-Lumi upgrade," *IEEE Trans. Appl. Supercond.*, vol. 31, no. 5, Aug. 2021, Art no. 4001804.

[18] G. Ambrosio et al., "Acceptance criteria part A: MQXFA magnet," FERMILAB-TM-2788-TD, Fermilab, Batavia, IL, USA, and CERN EDMS 2031083, 2020

[19] G. Ambrosio *et al*., "Challenges and Lessons Learned From Fabrication, Testing, and Analysis of Eight MQXFA Low Beta Quadrupole Magnets for HL-LHC," in *IEEE Transactions on Applied Superconductivity*, vol. 33, no. 5, pp. 1-8, Aug. 2023, Art no. 4003508, doi: 10.1109/TASC.2023.3261842.

[20] A. Moros et al., "A Metallurgical Inspection Method to Assess the Damage in Performance-Limiting $Nb_3Sn$ Accelerator Magnet Coils", IEEE Trans. Appl. Supercond. 33 (2023) 5, 4000208.

[21] S. I. Bermudez *et al*., "Performance of a MQXF $Nb_3Sn$ Quadrupole Magnet Under Different Stress Level," in *IEEE Transactions on Applied Superconductivity*, vol. 32, no. 6, pp. 1-6, Sept. 2022, 4007106.

[22] D. W. Cheng *et al*., "The Challenges and Solutions of Meeting the Assembly Specifications for the 4.5 m Long MQXFA Magnets for the Hi-Luminosity LHC," in *IEEE Transactions on Applied Superconductivity*, vol. 33, 5, 1-5, Aug. 2023, 4003905.

[23] S. Feher *et al*., "AUP First Pre-Series Cryo-Assembly Design Production and Test Overview," in *IEEE Transactions on Applied Superconductivity*, vol. 34, no. 5, pp. 1-5, Aug. 2024, Art no. 4005605, doi: 10.1109/TASC.2024.3379173.

[24] R. Bossert et al., "Design of the Fermilab pre-series cold mass for the HL-LHC accelerator upgrade project," *IEEE Trans. Appl. Supercond.*, vol. 32, no. 6, Sep. 2022, Art. no. 4002204, doi: 10.1109/TASC.2021.3060680

[25] A. Vouris et al., "Fabrication of the Fermilab pre-series cold mass for the HL-LHC accelerator upgrade project," *IEEE Trans. Appl. Supercond.*, vol. 31, no. 5, Aug. 2023, Art. no. 4002605, doi: 10.1109/TASC.2023.3254491.

[26] M. Baldini, S. Krave, R. Bossert, S. Feher, T. Strauss, and A. Vouris, "Application of distributed fiber optics strain sensors to LMQXFA cold mass welding," *IEEE Trans. Appl. Supercond.*, vol. 33, no. 5, Aug. 2023, Art. no. 9000705, doi: 10.1109/TASC.2023.3258904.

[27] R. Rabehl, S. Feher, D. Ramos, T. Strauss, and M. Struik, "AUP first pre-series cold mass installation into the cryostat," *IEEE Trans. Appl. Supercond.*, vol. 34, no. 5, Aug. 2024, Art. no. 4003705, doi: 10.1109/TASC.2024.3361871.

[28] G. Chlachidze et al., "Fermilab's horizontal test stand upgrade overview and commissioning," *IEEE Trans. Appl. Supercond.*, vol. 34, no. 5, Aug. 2024, Art. no. 9500104, doi: 10.1109/TASC.2023.3341985.

[29] S. Izquierdo Bermudez et al., "Overview of the Quench Heater Performance for MQXF, the $Nb_3Sn$ Low-β Quadrupole for the High Luminosity LHC," in *IEEE Trans. on Appl. Supercond.*, vol. 28, no. 4, pp. 1-6, June 2018, Art no. 4008406, doi: 10.1109/TASC.2018.2802839.

[30] Ravaioli, Emmanuele, "CLIQ. A new quench protection technology for superconducting magnets", PhD Thesis, Twente U., Enschede (2015).

[31] M. Baldini *et al*., "Quench Performance of the First Pre-Series AUP Cryo-Assembly," in *IEEE Transactions on Applied Superconductivity*, vol. 34, no. 5, pp. 1-4, Aug. 2024, Art no. 4005204, doi: 10.1109/TASC.2024.3358777.

[32] J. DiMarco et al., "Magnetic measurements and alignment results of LQXFA/B cold mass assemblies at Fermilab," *IEEE Trans. Appl. Supercond.*, vol. 34, no. 5, Aug. 2024, Art. No. 400205, doi: 10.1109/TASC.2023.3337202.